\documentclass[10pt]{iopart}

\usepackage{iopams}  
\usepackage{citesort}
\usepackage{graphics}
\usepackage{graphicx}

\newtheorem{prop}{\textbf{Proposition}}

\begin{document}
\title[Unbiased  diffusion of Brownian particles on disordered correlated potentials]{Unbiased  diffusion of Brownian particles on disordered correlated potentials}

\author{Ra\'ul Salgado-Garcia}
\ead{raulsg@uaem.mx}
\address{Centro de Investigaci\'on en Ciencias, Universidad Aut\'onoma del Estado de Morelos. Avenida Universidad 1001, colonia Chamilpa, C.P. 62209, Cuernavaca Morelos, Mexico.}

\author{Cesar Maldonado}

\address{Centro de Modelamiento Matem\'{a}tico, Universidad de Chile, Beauchef 851, Edificio Norte, Piso 7, Santiago, Chile.}

%
%

\begin{abstract}

In this work we study the diffusion of non-interacting overdamped particles, moving on  unbiased disordered correlated potentials, subjected to Gaussian white noise. We obtain an exact expression for the diffusion coefficient which allows us to prove that the unbiased diffusion of overdamped particles on a random polymer does not depend on the correlations of the disordered potentials. This universal behavior of the unbiased diffusivity is a direct consequence of the validity of the Einstein relation and the decay of correlations of the random polymer. We test the independence on correlations of the diffusion coefficient for correlated polymers produced by two different stochastic processes, a one-step Markov chain and the expansion-modification system. Within the accuracy of our simulations, we found that the numerically obtained diffusion coefficient for these systems  agree with the analytically calculated ones, confirming our predictions.
\end{abstract}

%
\vspace{2pc}

\submitto{\JSTAT}

\maketitle
%
%

\section{Introduction}

The diffusion of particles on one-dimensional (1D) random media has been the subject of intense research. Since the introduction of the Sinai's model it was recognized that disordered systems can exhibit normal and anomalous diffusion as well as normal and anomalous drift~\cite{sinai1983limiting}. Several models of diffusion in disordered media have been introduced since then~\cite{haus1987diffusion,havlin1987diffusion,bouchaud1990anomalous} in order to understand the underlying mechanisms leading to such behaviors. Particularly some simple deterministic models for transport in disordered potentials have been studied in order to understand the origin of the transport properties from the very deterministic dynamics~\cite{denisov2010biased,denisov2007analytically,denisov2007arrival,denisov2010anomalous,salgado2013normal}. A model for transport in random polymers that takes into account the pair correlations between monomers (to which we will refer to as the ``particle-polymer model'') has been introduced by the authors in Ref.~\cite{salgado2013normal}. In such a work it was shown that the deterministic biased diffusion is influenced by the presence of the correlations in the polymer. Indeed, in that paper, the authors gave an exact expression for the particle current and the diffusion coefficient in terms of the first two moments of the ``crossing times'' and its corresponding pair correlation function~\cite{salgado2013normal}, a result which is a consequence of the central limit theorem~\cite{gnedenko1968limit,gouezel2004central,chazottes2012fluctuations}. Based on the same arguments, it has been shown that when the particles along the polymer are placed at a finite temperature and subjected to a constant driving force, it is still possible to obtain an exact expression for the drift and diffusion coefficients in terms of the first and the second moments of the first passage  time (FPT) as well as its pair correlation function~\cite{salgado2014drift}. In this work we will show that, for the particle-polymer model,  the diffusion coefficient does not depend on the correlations between monomers along the polymer when the driving force is zero. The latter is a result that follows straightforwardly from the validity of the Einstein relation in disordered potentials at zero driving force. 

In consequence the paper is organized as follows. In Section~\ref{sec:model} we introduce the model as well as some general results concerning the transport process previously established. Then in Section~\ref{sec:zero-tilt} we give the exact expression for the diffusion coefficient in disordered polymers for zero driving force and introduce a simple model for the potential to illustrate the independence of the diffusivity on the correlations. First we use a Markov chain to produce exponentially correlated polymers which are used to induce a correlated potential. Then, we perform Langevin dynamics simulations to estimate the diffusion coefficient of overdamped particles moving on such polymers. We also perform analogous numerical experiments in the case in which the polymers are produced by means of the expansion-modification system, which is known to have polynomially decaying correlations.  Finally in Section~\ref{sec:conclusion} we give a brief discussion and the main conclusions of our work. Two appendices are given containing detailed calculations of our main results.

\section{The model}
\label{sec:model}

We will consider an ensemble of Brownian particles with overdamped dynamics moving on a 1D disordered potential $V(x)$ subjected to an external force $F$. The equation of motion of any of these particles is given by the stochastic differential equation,
\begin{equation}
\label{eq:overdamped_particle}
\gamma dX_t =  (f(X_t) +F)dt + \varrho_0 dW_t
\end{equation}
where $X_t$ stands for the position of the particle and $W_t$ is a standard Wiener process. The constants $F$ and $\gamma $ stand for the  driving force and the friction coefficient respectively. The constant  $\varrho_0$ represents the square root of the noise intensity (or the thermal fluctuations), which according to  the fluctuation-dissipation theorem satisfy the relation $\varrho_0^2 = 2 \gamma \beta^{-1}$.  Here $\beta$ denotes, as usual, the inverse temperature times the Boltzmann constant, $\beta = 1/k_B T$. 
The function $f(x)$ represents minus the gradient of the potential $V(x)$ that the particle feels due to its interaction with the substrate where the motion occurs.  As in Ref.~\cite{salgado2014drift}, we model the substrate (or the polymer) as a chain of ``unit cells''  of constant length $L$. The unit cells represent the monomers comprising the polymer. We denote by $\mathcal{A}$ the set of possible monomer types, a set that is assumed to be finite or countable infinite. The polymer is represented by a bi-infinite symbolic sequence  $\mathbf{a}:=(\dots,a_{-1},a_{0},a_{1},\dots)$, where $a_j \in \mathcal{A}$ stands for the monomer type located at the $j$-th cell, for all $j\in \mathbb{Z}$.  The set of possible random polymers will be denoted by $\mathcal{A}^{\mathbb{Z}}$ according to the conventional notation in symbolic dynamics~\cite{lind1995introduction}. As in Refs.~\cite{salgado2013normal,salgado2014drift}, we assume that the disordered potential $V(x)$ is the result of the interaction of a particle with the random polymer. 

Let $x\in \mathbb{R}$ denote the particle position along the substrate $\mathbf{a}\in\mathcal{A}^\mathbb{Z}$. A particle located at $x$ feels a (random) potential that is not only function the particle position $x$, but also of the substrate, i.e. $V(x) = \psi(x,\mathbf{a})$. This is a consequence of the fact that the particle might interact with the whole polymer and not only with the closest monomer. If $x=0$ we assume that the particle is located at the beginning of the $0$-th monomer $a_0$. In Ref.~\cite{salgado2014drift} it was shown that the random potential $ \psi(x,\mathbf{a})$ has the following property,
 \begin{equation}
\label{eq:shift-potential}
  \psi(x+nL,\mathbf{a})  = \psi[x,\sigma^n(\mathbf{a})].
\end{equation}
where $\sigma: \mathcal{A}^\mathbb{Z} \to \mathcal{A}^\mathbb{Z}$ is the \emph{shift mapping} defined as follows: if $\mathbf{a}, \mathbf{b}\in\mathcal{A}^\mathbb{Z} $ are such that $\sigma(\mathbf{a}) = \mathbf{b}$, then $b_i = a_{i+1}$ for all $i\in\mathbb{Z}$. As in Refs.~\cite{salgado2013normal,salgado2014drift} we assume that the substrate is generated by some stochastic process through a stationary  $\sigma$-invariant probability measure $\nu$ on $\mathcal{A}^\mathbb{Z}$. This hypothesis is equivalent to say that the statistical properties of the polymer are translationally invariant. 

In Ref.~\cite{salgado2014drift} it has been shown that the particle flux $J_{\mathrm{eff}}$ and the diffusion coefficient $D_{\mathrm{eff}}$ can be written exactly as follows,
\begin{eqnarray}
J_{\mathrm{eff}} &:=& \lim_{t \to \infty}\frac{ \langle\langle X_t \rangle\rangle}{t} = \frac{L}{\langle \langle \tau(0\to L) \rangle\rangle},
\label{eq:Jeff}
\\
D_{\mathrm{eff}} &:=& \lim_{t\to \infty} \frac{ \mathrm{Var} (X_t)}{ 2t} = 
\frac{ L^2 \varrho_\tau^2 }{2 \langle \langle \tau(0\to L) \rangle\rangle^3}.
\label{eq:Deff}
\end{eqnarray}
where $\tau(0\to L)$ is a random variable, called the \emph{first passage time} (FPT), defined as the time that the particle spends to reach, for the first time, the position $X_\tau = L$ from the initial condition $X_0 = 0$. The quantity $ \varrho_\tau^2$ is defined in terms of the variance of the FPT and its autocorrelation function as,
\begin{eqnarray}
 \varrho_\tau^2 &:=& \big\langle \big\langle \tau^2 (0\to L) \big\rangle\big\rangle - 
\big \langle\big \langle \tau(0\to L) \big\rangle\big\rangle^2 + 2 \sum_{m=1}^\infty C_\tau(m).
\nonumber
\end{eqnarray}
where
\begin{eqnarray}
C_\tau (m) &:=& \big\langle \big\langle \tau(0\to L) \tau\big(m L \to(m+1) L\big) \big\rangle\mathbf{\big\rangle} - 
\big \langle\big \langle \tau(0\to L) \big\rangle\big\rangle^2.
\nonumber
\end{eqnarray}
It is important to stress that the notation $\langle\langle \cdot \rangle\rangle$ denotes a double average, one with respect to the noise and the other with respect to the disorder. In Ref.~\cite{salgado2014drift} it was pointed out that is important to distinguish between these two different averages. If we put an ensemble of non-interacting  Brownian  particles over a (fixed) random polymer $\mathbf{a}$ we denote the average over the ensemble of particles as $\langle \cdot \rangle_{\mathrm{n}}$. This average will be referred to as the \emph{average with respect to the noise} (w.r.n.). Once a certain observable has been averaged with respect to the noise, it still depends on the specific realization of the polymer $\mathbf{a}$. Thus we need to perform a second average which should be carried out over an ensemble of different realizations of the random polymer. This average is taken with respect to a stationary measure $\nu$ which characterizes the process producing the random polymers. This average will be referred to as the \emph{average with respect to the polymers} (w.r.p.) and will be denoted by $\langle \cdot \rangle_{\mathrm{p}}$. Then, the double average $\langle\langle \cdot \rangle\rangle$ is actually achieved first by taking the average w.r.n.~and after that by taking the average w.r.p., i.e., $ \langle\langle \cdot \rangle\rangle = \langle\langle \cdot \rangle_\mathrm{n}\rangle_\mathrm{p}$. Additionally, we will use the following notation: $\mbox{Var}_\mathrm{n}(\mathcal{O})  := \langle \mathcal{O}^2 \rangle_{\mathrm{n}} - \langle \mathcal{O} \rangle_{\mathrm{n}}^2$  and $\mbox{Var}_\mathrm{p}(\mathcal{O}) := \langle \mathcal{O}^2 \rangle_{\mathrm{p}} - \langle \mathcal{O} \rangle_{\mathrm{p}}^2$  denote the variance of the observable $\mathcal{O}$ w.r.n.~and  w.r.p. respectively. Along this line, $\mbox{Var}(\mathcal{O})$ therefore denote the variance of the observable $\mathcal{O}$ with respect to both, the noise and the polymer ensemble, i.e., $\mbox{Var}(\mathcal{O}):=\langle \langle  \mathcal{O}^2  \rangle \rangle - \langle \langle \mathcal{O} \rangle \rangle^2$. 

Let us denote by $T_1(\mathbf{a})$ the mean FPT w.r.n., i.e., $T_1(\mathbf{a}) := \langle \tau(0\to L) \rangle_{\mathrm{n}}$. Then the drift and diffusion coefficients can be written as~\cite{salgado2014drift},
\begin{eqnarray}
\label{eq:drift}
J_{\mathrm{eff}} &=& \frac{L}{\langle T_1 \rangle_{\mathrm{p}}},
\\
D_{\mathrm{eff}} &=& D_{\mathrm{noisy}} + D_{\mathrm{det}},
\label{eq:diffusion}
\end{eqnarray}
where  $D_{\mathrm{noisy}}$ and $D_{\mathrm{det}}$ are referred to as the \emph{noisy and deterministic parts} of $D_{\mathrm{eff}}$ respectively. These quantities are defined as follows
\begin{eqnarray}
\label{eq:D_noisy}
D_{\mathrm{noisy}} &=&
\frac{L^2\big\langle \mathrm{Var}_{\mathrm{n}} (\tau( 0\to L)) \big \rangle_\mathrm{p} }{ 2 \langle T_1 \rangle_{\mathrm{p}}^3 },
\\
D_{\mathrm{det}} &=& \frac{ L^2 \mbox{Var}_{\mathrm{p}}(T_1) + 2 L^2 \sum_{m=1}^\infty C_\tau(m) }{2  \langle T_1 \rangle_{\mathrm{p}}^3}.
\label{eq:D_det}
\end{eqnarray}
The autocorrelation of the FPT can also be rewritten in terms of $T_1$, which gives
\begin{eqnarray}
\label{eq:correlation}
C_\tau (\ell) &=&
 \big\langle T_1(\mathbf{a}) T_1\left[ \sigma^{\ell}(\mathbf{a}) \right] \big\rangle_\mathrm{p} - 
\big \langle T_1 (\mathbf{a}) \big\rangle^2_\mathrm{p}.
\end{eqnarray}
The reason for expressing the transport coefficients, $J_{\mathrm{eff}}$ and $D_{\mathrm{eff}}$, in terms of the mean and the variance (with respect to the noise) of the FPT is that such quantities can be written exactly in terms of quadratures.  Indeed, in Ref.~\cite{salgado2014drift} it was proved that the mean FPT w.r.n., $T_1$, can be written as,
\begin{equation}
T_1(\mathbf{a}) = \gamma \beta \sum_{m=1}^\infty e^{-m\beta F L} q_{+}(\mathbf{a})q_{-}[\sigma^{-m}(\mathbf{a})] 
+ \gamma \beta I_0(\mathbf{a}),
\label{eq:T1}
\end{equation}
and the variance of the FPT, $\mathrm{Var}_{\mathrm{n}} [\tau( 0\to L)]$, as
\begin{eqnarray}
 \fl
\mathrm{Var}_{\mathrm{n}} [\tau( 0\to L)] &=& \,
 2(\gamma\beta)^2 \Bigg\{
\sum_{n=1}^\infty
\sum_{m=1}^\infty 
\sum_{l=1}^\infty
\bigg(  e^{-(n+m+l)\beta F L} q_{+}(\mathbf{a}) q_{+}[\sigma^{-n}(\mathbf{a})] 
\nonumber
\\
&\times& \,
q_{-}[\sigma^{-m-n}(\mathbf{a})]q_{-}[\sigma^{-n-l}(\mathbf{a})] \bigg)
\nonumber
\\
&+&
2  \sum_{n=1}^\infty \sum_{m=1}^\infty e^{-(n+m)\beta F L}
q_{+}(\mathbf{a}) q_{-}[\sigma^{-n-m}(\mathbf{a})] I_0[\sigma^{-n}(\mathbf{a})]
\nonumber
\\
&+& \sum_{n=1}^\infty e^{-n\beta F L}
q_{+}(\mathbf{a})I_1[\sigma^{-n}(\mathbf{a})]
\nonumber
\\
&+&
  \sum_{m=1}^\infty \sum_{l=1}^\infty e^{-(m+l)\beta F L}
q_{-}[\sigma^{-m}(\mathbf{a})] q_{-}[\sigma^{-l}(\mathbf{a})] 
 I_2(\mathbf{a})
\nonumber
\\
&+&
 2   \sum_{m=1}^\infty e^{-m\beta F L}
q_{-}[\sigma^{-m}(\mathbf{a})]
 I_3(\mathbf{a})
+ I_4(\mathbf{a}) \Bigg\}
\end{eqnarray}
which defines completely the transport coefficients. Here, the functions $q_+,q_- :\mathcal{A}^\mathbb{Z} \to \mathbb{R}$ were defined as
\begin{eqnarray}
\label{eq:def-qpm}
q_\pm (\mathbf{a} ) &:=& \int_0^L dx \exp\big( \pm \beta [ \psi(x,\mathbf{a}) - xF] \big),
\end{eqnarray}
and the functions $B_{\pm} : \mathbb{R} \times\mathcal{A}^\mathbb{Z} \to \mathbb{R} $ and $Q_{\pm} :  \mathbb{R} \times\mathcal{A}^\mathbb{Z} \to \mathbb{R} $ were defined as,
\begin{eqnarray}
\label{eq:def-Qpm}
Q_\pm (x,\mathbf{a}) &:=& \int_0^x dy \exp\big( \pm \beta [ \psi(y,\mathbf{a}) - yF] \big),
\\
\label{eq_def-Bpm}
B_\pm (x,\mathbf{a}) &:=&\exp\big( \pm \beta [ \psi(x,\mathbf{a}) - xF] \big).
\end{eqnarray}
We also used the following short-hand notation for the involved integrals,
\begin{eqnarray}
I_0 (\mathbf{a}) &=&  \int_0^L Q_-(x,\mathbf{a}) B_+(x,\mathbf{a})dx,
\nonumber
\\
I_1(\mathbf{a}) &=& \int_0^L \left[Q_-(x,\mathbf{a})\right]^2 B_+(x,\mathbf{a})dx,
\nonumber
\\
I_2(\mathbf{a}) &=&\int_0^L Q_+(x,\mathbf{a}) B_+(x,\mathbf{a})dx,
\nonumber
\\
I_3(\mathbf{a})&=&\int_0^L B_+(x,\mathbf{a}) \int_0^x Q_-(u,\mathbf{a}) B_+(u,\mathbf{a})dudx,
\nonumber
\\
I_4(\mathbf{a})&=&\int_0^L B_+(x,\mathbf{a}) \int_0^x B_+(u,\mathbf{a}) \left[Q_-(u,\mathbf{a}) \right]^2 dudx.
\label{eq:integrals}
\end{eqnarray}

In Ref.~\cite{salgado2014drift} it was shown that, for the case of non-correlated random polymers, the transport of particles in this system arises an interesting phenomenon which is the non-monotonic behavior of the diffusivity on the temperature. In this work we will study the effective diffusion coefficient of overdamped particles moving on correlated and uncorrelated potentials. Bellow we will show with certain generality that the unbiased diffusion does not depend on the correlations of the polymer beyond the number of monomers that the particle interacts with.

\section{Unbiased diffusion on random polymers}
\label{sec:zero-tilt}

A remarkable property that can be inferred from our formalism, is that the diffusivity in unbiased disordered potentials does not depend on the correlations of the substrate. This means that the particles in disordered potentials are unable to feel the pair correlations between monomers comprising the polymer. Consequently, the transport properties of a system of particles diffusing on correlated polymers are equal to those of a system of particles moving on an ``equivalent'' ensemble of uncorrelated polymers. Recently~\cite{PhysRevLett.113.100601} this kind of behavior has been shown to occur in a similar model of diffusion on random potentials. In Ref.~\cite{PhysRevLett.113.100601} Goychuk and Kharchenko calculated the diffusion coefficient  of overdamped particles diffusing on a correlated random potential with a Gaussian distribution. Surprisingly their result does not depend on the correlations, a conclusion that is reached by using the Einstein relation. Here we prove that for our model of random potentials the diffusion coefficient is independent on the correlations if the substrate is \emph{ergodic} and has decay of correlations. The word ergodic in this context means that the polymer average can be taken as a time average if we think of the polymer as a discrete-time stochastic process. 

Bouchaud and Georges in Ref.~\cite{bouchaud1990anomalous}  have shown that the Einstein relation, $\mu =\beta D_{\mathrm{eff}}(F=0) $, remains valid for the class of disordered potentials we are considering. The constant $\mu$ stands for the mobility at zero bias, $\mu = \lim_{F\to 0} J(F)/F$. From the Einstein relation it is easy to obtain the diffusion coefficient since
\[
D_{\mathrm{eff}} (F = 0) = \frac{1}{\beta}\lim_{F\to 0}\frac{J_{\mathrm{eff}} (F)}{F}.
\]
 In ~\ref{ape:Deff_zero_tilt} we prove that the first moment of the FPT the divergence occurs as $F^{-1}$. This implies that the particle current vanishes, as expected, in the limit $F\to 0$. Actually the particle current behaves linearly in a small neighborhood of $F=0$.  This allows us to obtain the effective diffusion coefficient as follows
\begin{equation}
D_{\mathrm{eff}} (F = 0)  = \frac{L^2 }{\gamma \beta Z_+ Z_-}.
\label{eq:Deff_F=0}
\end{equation}
Here the constants $Z_\pm$ are defined as
\begin{equation}
Z_\pm :=  \langle q_\pm (\mathbf{a}) \rangle_{\mathrm{p}} =  \left \langle \int_0^L \exp{ \left[ \pm\beta \psi(x, \mathbf{a}) \right]} \, dx \right\rangle_{\mathrm{p}}.
\label{eq:Zpm}
\end{equation}

Notice that the effective diffusion coefficient depends only on the quantities $Z_\pm$ which are \emph{local averages}, in the sense that the polymers involved in the averages are not shifted with respect to each other (as it occurs in the correlation function given in Eq.~(\ref{eq:correlation})). Thus, the expression~(\ref{eq:Deff_F=0}) for the effective diffusion coefficient implies that, on \emph{ergodic} polymers with decay of correlations, the diffusivity does not depend on the correlations between monomers at distant sites if the particle only interacts with the closest monomer to it (i.e., if the potential $\psi (x,\mathbf{a})$ only depends on one ``coordinate'' of $\mathbf{a}$). This means that particles on polymers with long range correlations exhibit the same diffusivity as the particles that moves on polymers without correlations but with the same one-dimensional marginals. More generally, the diffusivity of a particle moving on a random polymer will be influenced by the pair correlations between monomers as distant as the \emph{range} of the potential induced by the interaction. Here, the \emph{range} of a potential stands for the maximal distance between two monomers having a non vanishing interaction with the particle. This makes the averages with respect to the polymer ensemble to depend on the range of the interaction. In other words, if the range of the potential $\psi (x,\mathbf{a})$ is $n$, then, $\psi$ can be seen as a function of $n$ coordinates of $\mathbf{a}$, i.e., $\psi (x,\mathbf{a}) = \psi (y,a_j, a_{j+1}, \dots, a_{j+n-1})$. Therefore, to perform the polymer averages $Z_\pm$ it is necessary to have at least the $n$-dimensional marginals of the probability measure $\nu$ defining the polymer ensemble i.e., we need to know the joint probability distribution of $n$ monomers, $\mathbb{P}(a_j,a_{j+1},\dots,a_{j+n-1})$. With exception of the Bernoulli measure, the $n$-marginals cannot be decomposed, in general, as the product of the $1$-dimensional marginals. In the following we give some examples to better illustrate these situations.

\subsection{Potential of range $n=1$}

\begin{figure}[h]
\begin{center}
\scalebox{0.35}{\includegraphics{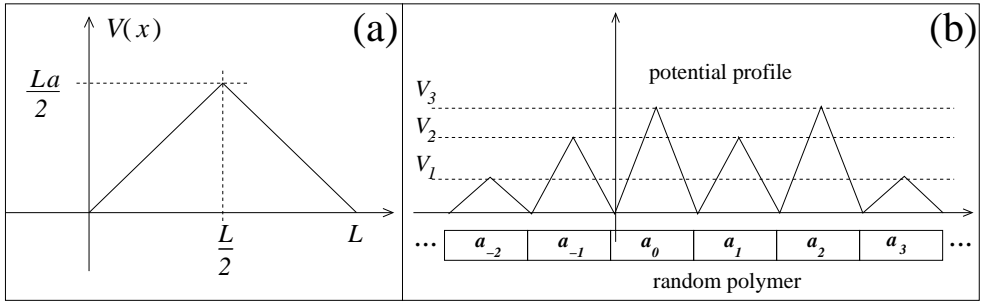}}
\end{center}
     \caption{
     Schematic representation of the potential model. (a) The potential profile on the $0$th unit cell. (b) A realization of the random potential with a few unit cells. In this case every monomer along the chain can be taken among three possible monomers ($k = 3$) with heights $V_1 = f_1 L/2$,   $V_2 = f_2 L/2$ and $V_3 = f_3 L/2$. To perform analytical calculations as well as numerical simulations we have taken the values $f_1 = 0.8$, $f_2 = 4.2$, and  $f_3 = 9.0$  (see text).
          }
\label{fig:potential-model}
\end{figure}

In order to exemplify the independence on the correlations of the diffusion coefficient at zero tilt strength we use the potential model introduced in Ref.~\cite{salgado2014drift}. The potential model $V(x)$ used in such a work is of range $1$ since the potential only depends on the monomer at which the particle is located. Explicitly the potential is defined as follows. If $x$ represents the particle position and we write as $x = y + nL$, then
\begin{equation}
V(x)= \psi (x,a_n) =  \left\{ \begin{array} 
            {r@{\quad \mbox{ if } \quad}l} 
   a_n y   &  0\leq  y<L/2    \\ 
   a_n (L-y )  &  L/2 \leq y < L.       \\ 
             \end{array} \right. 
\label{eq:potential-model}
\end{equation}
Here $n\in \mathbb{Z}$ stands for the index of the unit cell at which the particle is located and $y$ is the relative position of the particle along such unit cell.  The varible $a_n$ is a random variable representing the monomer type at the $n$th unit cell. This potential model is shown schematically in Fig.~\ref{fig:potential-model}. We observe that such potential is symmetric on every unit cell with a maximum located at $y=1/2$. The height of the potential assumed to be random taking values from a finite set. Since the height of the potential is given by $a_n L/2$ we can assume that $a_n$ represents the random variable, for every $n\in\mathbb{Z}$,  which can take values from a set $\mathcal{A} :=\{ f_j \, :\,  1\leq j \leq k \}$.
%

We should stress here that the sequence $\mathbf{a} := (\dots, a_{-1}, a_0, a_1,\dots) \in \mathcal{A}^\mathbb{Z} $ represents the polymer (with $a_n$ the corresponding monomers for $n\in\mathbb{Z}$). The values $f_j$  represent the ``slopes'' that the potential can take and, in some way, stand for the  possible monomer types from which the polymer is built up. It is clear that the proposed potential depends only on one monomer, i.e., $V(x) = \psi(y, a_n)$ if $x = y+nL$. Since we are considering the Bernoulli measure on $\mathcal{A}^\mathbb{Z}$, we only need to specify the probability that a given monomer $a_n$  equals a monomer type $f_j$ for $1\leq j \leq k$. In other words, we need to specify the $1$-dimensional marginal, i.e.,  $\nu(a_n=f_j) =: p(f_j)$ to perform the polymer average.   With these quantities we can state explicitly how to take the average with respect to the polymer ensemble. If $h : \mathcal{A}  \to \mathbb{R}$, then we have
\begin{equation}
\label{eq:polymer_avg}
\langle h(a) \rangle_{\mathrm{p}} = \sum_{j=1}^k h(f_j) p(f_j).
\end{equation}
Notice that this average does not depend on $n$, which reflects the fact that the chosen measure is translationally invariant (or shift-invariant). With this potential model we have that the integrals $Z_+$ and  $Z_-$ can be done exactly, giving
\begin{eqnarray}
Z_\pm &=& 2 \left \langle \frac{\exp\left( \pm \beta a L/2\right) -1 }{\pm \beta a} \right \rangle_{\mathrm{p}} 
\nonumber 
\\
&=&
 2 \sum_{j=1}^k 
\frac{\exp\left( \pm \beta f_j L/2\right) -1 }{\pm \beta f_j}
p(f_j).
\end{eqnarray}
The presence of correlations does not change the above average. This is because the function to be averaged only depends on one coordinate of $\mathbf{a}\in\mathcal{A}^\mathbb{Z}$. Thus, to perform the average over the polymer ensemble we only require the $1$-dimensional marginals. Thus if two stochastic processes producing the (random) substrate have identical $1$-dimensional marginals then they will exhibit the same diffusion coefficient at all temperatures. To confirm that it is the case, we simulate the Langevin dynamics of $10\,000$ overdamped particles diffusing on several random polymers. Then we calculate the diffusion coefficient from the exact expression given in Eq.~(\ref{eq:Deff_F=0}) and compare it with the diffusivity obtained from the simulations. To perform the referred numerical simulations first we built the polymers using a set of $k = 3$ monomer types.  The slopes defining the corresponding monomer types  are chosen as $f_1 = 0.8$, $f_2 = 4.2$ and $f_3  = 9$. The set of possible monomer types will be represented by $\mathcal{A} := \{1,2,3\}$. Thus, a realization of the polymer $\mathbf{a}  = (\dots,a_{-1},a_0,a_1,\dots)$ has an associated sequence of potential profiles with ``slopes'' $(\dots,f_{a_{-1}},f_{a_0},f_{a_1},\dots)$. In our numerical simulations we only vary the way in which the polymer is built up. For numerical reasons we fix the  parameters $L$ and $\gamma $ to one throughout the rest of this work. 

\begin{figure}[h]
\begin{center}
\scalebox{0.4}{\includegraphics{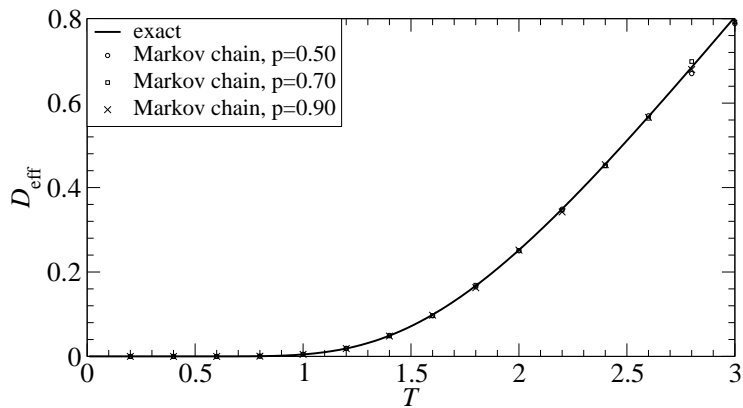}}
\end{center}
     \caption{
     Effective diffusion coefficient for overdamped particles on ``Markov''  and ``Bernoulli'' polymers. We simulate the dynamics of $10000$ overdamped particles on random polymers built up by means of a Markov chain. The stochastic matrix corresponding to such a Markov process is given in Eq.~(\ref{eq:stochastic_matrix}) having a parameter $p$. We calculated the diffusion coefficient from the simulations for three different cases $p=0.50$, $p=0.70$ and $p=0.90$ of the Markov chain. We also calculate the diffusion coefficient for the case of the Bernoulli process. We observe that the diffusion coefficients for the Markov chain do not vary with the parameter $p$ and  coincide with the diffusion coefficient obtained from the Bernoulli process. Moreover, these diffusion coefficients are consistent with the values predicted by Eq.~(\ref{eq:Deff_F=0}). This gives an example of the fact that  the  diffusion coefficient (in absence of driving force) is independent on the correlations present in the polymers.
          }
\label{fig:Deff_MarkovChain_F=0}
\end{figure}

In Fig.~\ref{fig:Deff_MarkovChain_F=0} we observe the diffusion coefficient, as a function of the temperature, for particles moving on random polymers built up by means of a Bernoulli process and a Markov chain. On one hand, the Bernoulli process is defined uniquely by the 1-dimensional marginals, $p_j = \nu(a_n = f_j) $ for $j=1,2,3$. Notice that the latter does no depend on the lattice site $n$, which means that the measure is shift invariant. For our numerical simulations we use the equidistribution for the monomer types, i.e.,  $p_1 = p_2 = p_3 = 1/3$. This means that the polymer is built up at random as follows: at each unit cell we chose at random a monomer type from the set $\mathcal{A} := \{1,2,3\}$ with the same probability. Such a choice is independent of the monomers chosen in the rest of the chain. On the other hand, the Markov process that we use to built up the polymers is defined through the stochastic matrix $P : \mathcal{A}\times \mathcal{A} \to [0,1]$ given by
\begin{equation}
P = \left(
  \begin{array}{ccc}
   0 & p & q \\
   q & 0 & p \\
   p & q & 0
  \end{array} \right),
  \label{eq:stochastic_matrix}
\end{equation}
where $p$ and $q$ are parameters such that $q=1-p$. It is easy to see that this matrix is doubly stochastic and the unique invariant probability vector $\mathbf{\pi} = \mathbf{\pi} P $ is given by $\pi = (\frac{1}{3},\frac{1}{3},\frac{1}{3})$. It is known that the correlations for any Markov process decay exponentially fast~\cite{levin2009markov}. Contrary to the Bernoulli process, the monomer type that is chosen at random at a given cell depends on the choice already made at an adjacent  unit cell. 
%
Notice that although the Bernoulli and the Markov processes are different in nature as explained above, they have the same 1-dimensional marginals. This means that the probability that a given monomer type appears along the polymer is the same regardless the process (Markovian or Bernoulli). This is a consequence of the fact the components of the invariant probability vector  $\pi = (1/3,1/3,1/3) $ of the stochastic matrix $P$, actually correspond to the 1-dimensional marginals of the stationary Markov chain.  In Fig.~\ref{fig:Deff_MarkovChain_F=0} we can appreciate that,  within the accuracy of our numerical simulations, the diffusion coefficient for different realizations of the polymers  coincide,  regardless if they are built up by means of the stationary Markov chain or by means of the Bernoulli process. Moreover, these numerical experiments fits satisfactorily (within the accuracy of our simulations) to the exact curve calculated from Eq.~(\ref{eq:Deff_F=0}).

\subsection{Potentials of range $n=2$}

In order to test the independence on long-range correlations of the diffusion coefficient we simulate the dynamics of overdamped particles on random polymers built up by means of the expansion modification (E-M) process. The E-M system is a stochastic process that was introduced as a simple model exhibiting spatial $1/f$ noise~\cite{li1989spatial,li1991expansion} and, in particular, it was used to understand the long-range correlation present in DNA sequences~\cite{li1992long,li1994understanding}. Such a system is defined by means of two fundamental processes as follows. Consider a ``seed'' (a symbol in the  binary alphabet $\{0,1\}$). The seed is  expanded with a probability $p$ and it is modified (it is replaced by the complementary symbol) with probability $1-p$. In other words, if $a\in \{0,1\}$ then, under the E-M process this symbol is subjected to the transformation,
\[
a \mapsto \left\{\begin{array}{cl} \overline{a} &  \mbox{ with probability } 1-p \\ 
                                   a a & \mbox{ with probability } p.
                  \end{array}\right.
\]
and these transformations are extended coordinate-wise to words of arbitrary length in order to define the infinitely iterated E-M process. Here $\overline{a}$ stands for the complementary symbol, i.e., $\overline{1} = 0$ and $\overline{0} = 1$.
This process allows the seed to grow in such a way that the resulting string of symbols becomes infinite with probability one.  This stochastic process has been studied within the context of symbolic dynamics in Ref.~\cite{salgado2013exact}. In that work it was rigorously proved that the E-M system has a unique stationary measure having polynomial decay of correlations for an open set of values of the parameter $p$. Specifically, they proved that the pair correlation function $C(\ell)$ has a behavior of the form $C(\ell) \asymp \ell^{\theta}$ where $\theta$ is an exponent depending on the parameter $p$, i.e. $ \theta = \theta( p)$.  The authors of Ref.~\cite{salgado2013exact} also conjectured that this behavior for the correlation functions occurs for almost all the values of the expansion probability and gave an explicit formula for the  corresponding exponent $\theta(p)$,
\begin{equation}
\label{eq:exponent_EM}
\theta (p) = \frac{ \log(1+p)-\log(|2p-1|)-\log(|3p -1|)}{ \log(1+p)}.
\end{equation}
The treatment of the E-M system by means of symbolic dynamics allows us to obtain exact formulas for the $n$-dimensional marginals defining the ($\sigma$-invariant) stationary probability measure~\cite{salgado2013exact}.   Indeed, the $n$-dimensional marginals are obtained as the stationary probability vectors of a hierarchy of stochastic matrices of size $2^n \times 2^n$. Every stochastic matrix in such a hierarchy gives the ``transition probability'' from one word of size $n$ to another one under the action of expansions and modifications on the polymer. Following the treatment of Ref.~\cite{salgado2013exact}, we prove  in~\ref{ape:marginals_EM} that the stochastic matrices for the cases $n=1$ and $n=2$ are given by,
\begin{eqnarray}
M_1 &=& \frac{1}{1+p} \left(
  \begin{array}{cc}
   2p & 1-p  \\
   1-p & 2p   
   \end{array} \right),
  \label{eq:M1}
\\
M_2 &=& \frac{1}{1+p} \left(
  \begin{array}{cccc}
   2p^2+pq  & pq   & pq   & q^2 \\
   p^2 + 2pq & p^2  & q^2  & pq  \\
   pq       & q^2  & p^2  & p^2 +2pq \\
   q^2      & pq   & pq   & 2p^2+pq     
  \end{array} \right).
  \label{eq:M2}
\end{eqnarray}
The respective invariant probability vectors $v_1 :\mathcal{A} \to [0,1]$ and $v_2: \mathcal{A}^2 \to [0,1]$ are given by
\begin{eqnarray}
v_1 &=& \left(\frac{1}{2},\frac{1}{2}\right),
\label{eq:v1}
\\
v_2 &=& \left( \frac{3-2p}{10-8p}, \frac{1-p}{5-4p}, \frac{1-p}{5-4p} ,\frac{3-2p}{10-8p} \right),
\label{eq:v2}
\end{eqnarray}
which are obtained as the left eigenvectors associated to the largest eigenvalues of $M_1$ and $M_2$ respectively. 
With these quantities we can evaluate the average w.r.p. of any function on $\mathcal{A}^\mathbb{Z}$  that depends on one or two coordinates of the polymer. Specifically, if we have a function depending on one coordinate of the polymer,  $g : \mathcal{A} \to \mathbb{R}$, then the average w.r.p. is defined by
\begin{equation}
\langle g (a)\rangle_\mathrm{p} = \sum_{a=0}^1 g(a) v_1(a).
\end{equation}  
On the other hand if we have a function depending on two coordinates of the polymer, $h : \mathcal{A}^2 \to \mathbb{R}$, then the average w.r.p. should be achieved  as follows,
\begin{equation}
\langle h (a,b)\rangle_\mathrm{p} = \sum_{a=0}^1\sum_{b=0}^1 h(a,b) v_2(a,b),
\end{equation}  
a summation that will be represented alternatively as
\begin{equation}
\langle h (\mathbf{a})\rangle_\mathrm{p} = \sum_{\mathbf{a}\in{\mathcal{A}^2}} h(\mathbf{a}) v_2(\mathbf{a}).
\end{equation}  

Now let us define a toy model for the potential induced by the particle-polymer interaction.  We will assume that the particle interacts with two monomers when it is placed on the polymer. This interaction potential will depend on two coordinates of the random polymer. We  will assume additionally that the induced potential $\psi$ is piece-wise linear. Given a particle placed at the $n$th unit cell  on the polymer $\mathbf{a}$, we write the particle position $x$ as $x = y +nL$ with $y\in [0,1)$. Remember that in this case the polymer is built up from a binary ``alphabet'' $\mathcal{A} = \{0,1\}$ (since we have only two symbols in the expansion-modification process). Then the potential is defined as
\begin{equation}
\fl
\qquad
\qquad
\psi(x, a_n,a_{n+1}) =  \left\{ \begin{array} 
            {r@{\quad \mbox{ if } \quad}l} 
  \varphi( a_n , a_{n+1}) y   &  0\leq  y<L/2    \\ 
   \varphi( a_n , a_{n+1}) (L-y )  &  L/2 \leq y < L.     \\ 
             \end{array} \right. 
\label{eq:potential-model_2}
\end{equation}
%
%
\begin{figure}[t]
\begin{center}
\scalebox{0.4}{\includegraphics{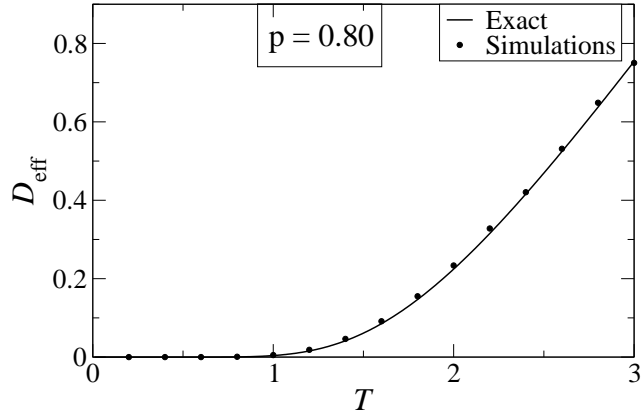}}
\end{center}
     \caption{ Effective diffusion coefficient of overdamped particles moving on E-M polymers with expansion probability $p=0.80$. In this figure we display the effective diffusion coefficient estimated from the Langevin dynamic simulations (filled circles) and the theoretical prediction from Eq.~(\ref{eq:Deff_F=0}) (solid line). This figure clearly shows that the effective diffusion coefficient obtained from the simulations is consistent with the theoretical prediction, at least within the accuracy of our numerical experiments.  
                 }
\label{fig:Deff-p=0.80}
\end{figure}
%
%
The function $\varphi : \mathcal{A}^2 \times\mathcal{A}^2 \to \mathbb{R}$ appearing in the equation above is such that,
\begin{equation}
\fl
\varphi(a,b) =  \left\{ \begin{array} 
            {r@{\quad \mbox{ if } \quad}l} 
  f_1   &  a = 0 \, \mbox{ and } \, b = 0    \\ 
  f_2  &   \left( a = 0 \, \mbox{ and } \, b = 1 \right)  \quad \mbox{or} \quad   \left(  a = 1 \, \mbox{ and } \, b = 0 \right)  \\ 
  f_3  &    a = 1 \, \mbox{ and } \, b = 1.     \\ 
             \end{array} \right. 
\label{eq:def_varphi}
\end{equation}
The values $f_i$ for $i = 1,2,3$ are chosen as in the preceding example, $f_1 = 0.80$, $f_2 = 4.2$ and $f_3 = 9.0$. With this model we have that the resulting potential on a given unit cell depends not only on the monomer type located in such a unit cell but also depends on the right adjacent monomer type. Note that the potential profile defined by~(\ref{eq:potential-model_2}) is the analogous to the potential profile defined by Eq.~(\ref{eq:potential-model}). The difference is that in the present model the slopes are defined by two coordinates of the polymer, instead of one as in the potential model of Eq.~(\ref{eq:potential-model}). With these definition we can now evaluate the quantities $Z_+$ and $Z_-$ defined in Eq.~(\ref{eq:Zpm}) which allows us to obtain an expression for the diffusion coefficient through  Eq.~(\ref{eq:Deff_F=0}). Notice that the integral appearing in Eq.~(\ref{eq:Zpm}) is the same as in the case of potentials of range $n=1$. Thus, it is easy to see that,
\begin{eqnarray}
Z_\pm &=& 2 \left \langle \frac{\exp\left( \pm \beta \varphi(\mathbf{a}) L/2\right) -1 }{\pm \beta \varphi(\mathbf{a})} \right \rangle_{\mathrm{p}} 
\nonumber 
\\
&=&
 2 \sum_{\mathbf{a} \in \mathcal{A}^2} \bigg[
\frac{\exp\left( \pm \beta \varphi(\mathbf{a})  L/2\right) -1 }{\pm \beta \varphi(\mathbf{a}) }\bigg]
v_2(\mathbf{a}).
\end{eqnarray}

In Fig.~\ref{fig:Deff-p=0.80} we show the effective diffusion coefficient for particles along an E-M polymer with expansion probability $p=0.80$. In this case the pair correlation function of the polymer is approximately given by $C(\ell) \approx \ell^{\theta(0.80)}$, where $\theta(0.80) \approx 1.296$, which implies that the correlations decay too slowly. To perform the numerical experiment (from which we estimate the effective diffusion coefficient) we simulated the Langevin dynamics of $35\,000$ particles, every particle moving on different realizations of the polymer, during a total time of $10^5$ arb.~units. We can appreciate in Fig.~\ref{fig:Deff-p=0.80}  that even in this situation, where the polymer has long range correlations, the diffusion coefficient obtained from the Langevin dynamics (filled circles) fits satisfactorily (within the accuracy of our simulations) with the diffusion coefficient predicted by Eq.~(\ref{eq:Deff_F=0}) (solid line). This means, again, that the diffusivity does not depend on the large correlation present in the disordered substrate.

\begin{figure}[t]
\begin{center}
\scalebox{0.4}{\includegraphics{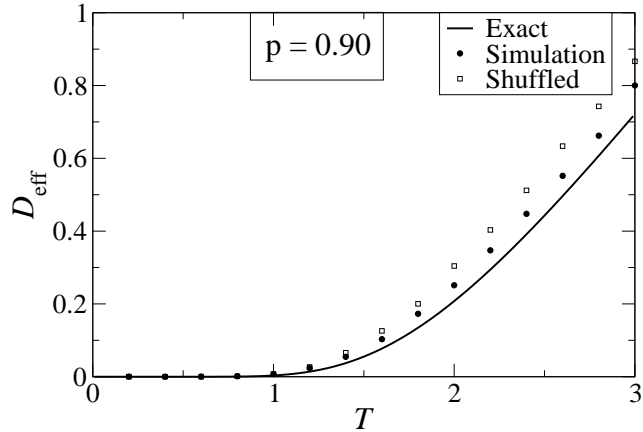}}
\end{center}
     \caption{
Effective diffusion coefficient of overdamped particles moving on E-M polymers with expansion probability $p=0.90$. In this figure we display the effective diffusion coefficient estimated from the Langevin dynamic simulations (filled circles) and the theoretical prediction from Eq.~(\ref{eq:Deff_F=0}) (solid line). We should notice a clear (non-random) deviation of the simulations with respect to the theoretical curve. In this figure we also display the effective diffusion coefficient of particle on the polymers produced by the E-M process but after shuffling the symbols. The shuffling process ``destroy'' the correlations and, as we can see, this diffusion coefficient differs from both, the theoretical result and the diffusivity originally calculated in the correlated polymer. This result evidence the fact that the E-M process has not reached its stationary state and the fact that the diffusion process in long-range correlated polymers undergoes a long transient resulting possible in a very slow convergence of the diffusion coefficient.  
                 }
\label{fig:Deff-p=0.90}
\end{figure}
%

\begin{figure}[t]
\begin{center}
\scalebox{0.4}{\includegraphics{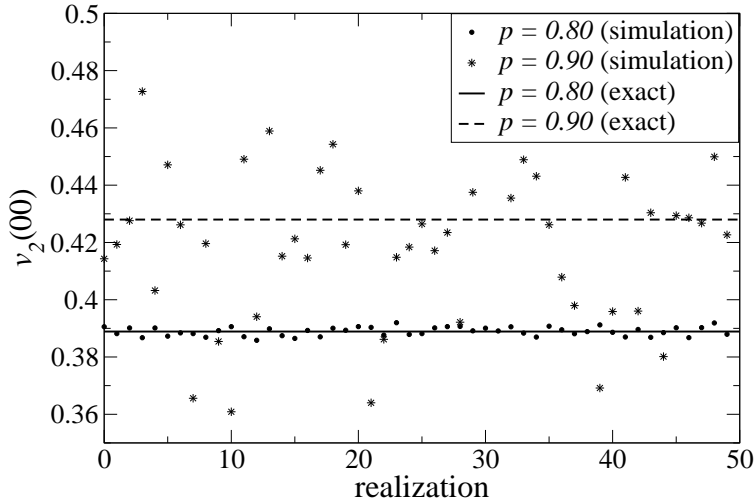}}
\end{center}
     \caption{
Two-dimensional marginal of the polymer corresponding to the word $\mathrm{00}$. We display the relative frequency of occurrence  of the word $\mathrm{00}$ along every realization of the polymer. Every realization has $10^6$ symbols and we produced $50$ different realizations for the cases $p=0.80$ and $p=0.90$. We observe that in the case $p=0.80$ the numerically obtained two-dimensional marginal $v_2(00)$  is consistent with the predicted by Eq.~(\ref{eq:v2}). We also appreciate that for the case $p=0.90$ we have large statistical deviation from the theoretical value compared to the case $p=0.90$. Actually, the standard deviation for $p=0.90$ is an order of magnitude larger than the standard deviation of the case $p=0.80$.  
                 }
\label{fig:v2_00}
\end{figure}

We now test our formula~(\ref{eq:Deff_F=0}) for the effective diffusivity in a more dramatic situation. This is the case where the correlations  decay much more slowly in such a way that the correlation function behaves asymptotically as $C(\ell) \approx \ell^{\theta(0.90)}$ where $\theta (0.90) \approx 0.521$. In this case, a typical realization of the E-M polymer has large ordered domains which are responsible of the large correlations present in the (disordered) substrate. It is natural to expect, in the present case, large deviated results from the theoretical ones. This is, effectively, the case as we observe in Fig.~\ref{fig:Deff-p=0.90}. The effective diffusion coefficient obtained from Langevin dynamic simulations (filled circles) deviates moderately (but with a clear non random tendency) from the theoretical curve predicted by Eq.~(\ref{eq:Deff_F=0}) (solid line). This behavior could be the result of two phenomena naturally present in the system. First we should remind that the E-M process relaxes slowly to its stationary state, especially for large values of the expansion probability $p$~\cite{salgado2013exact}. As it was stated in Ref.~\cite{salgado2013exact}, we should be aware of this phenomenon since it could lead to distorted  observations. On the other hand, another phenomenon that could also be responsible of the deviations observed in Fig.~\ref{fig:Deff-p=0.90} is the fact that the diffusion of the particles is by itself a very slow process. Indeed, the effective diffusivity is a property of the whole polymer, and if the particles are unable to visit the whole polymer, the diffusion coefficient would have large statistical errors. To verify that these two phenomena are actually present in the numerical simulations we performed the following experiment. We toke the polymer produced by the E-M process (with $p=0.90$) and we shuffled its monomers in order to ``destroy'' the large correlations on the substrate. This procedure was done by taking two monomers at random and interchanging its position along the polymer. We performed $10^6$ ``interchanges'' on every polymer of $10^6$ symbols long.  Then we placed the particles on the shuffled polymers and performed the corresponding Langevin simulations. In Fig.~\ref{fig:Deff-p=0.90} we display the diffusion coefficient calculated in this way (open squares). We appreciate that the resulting diffusion coefficient still deviates from the theoretical one. This means that the deviation of the diffusivity observed in the long range correlated polymers does not come from the correlations at all, but from the fact that the polymer itself, has not reached its stationary state. This phenomenon can be appreciated in Fig.~\ref{fig:v2_00}. In such a figure we plot the relative frequency of occurrence of the word $\mathrm{00}$  for $50$ realizations of the random polymer of $10^6$ symbols long. We can observe that for the case $p=0.80$ such relative frequency (filled circles) coincide in the average with the theoretical value (solid line), the two-dimensional marginal $v_2(00)$,  predicted by Eq.~(\ref{eq:v2}). On the other hand, for the case $p=0.90$, we can appreciate that the relative frequencies obtained from the simulations (stars) deviate largely from the theoretical prediction (dashed line) compared to the case $p=0.80$. Indeed,  the standard deviation of the relative frequency for the case $p=0.80$  is $\Delta v \approx 0.001550471$ while for the case $p=0.90$ we obtain  $\Delta v \approx  0.02874428$ which differs in one order of magnitude. This behavior clearly affects the effective diffusion coefficient obtained from the simulations even in the case where the correlations are destroyed by shuffling the symbols. 

%
%
%

\section{Discussion and conclusions}
\label{sec:conclusion}

In this work we have considered a model for unbiased diffusion of particles moving along disordered correlated polymers. We have proved that if the particle only interacts with the monomer closest to it, the diffusivity does not depend on the correlations between monomers along the polymer. In general we have proved that if the particle interacts with a few monomers on the polymer, it will depend only on the correlations of monomers as distant as the range of the interaction. This result was proved under the assumptions ($i$) that the Einstein relation is valid in this kind of systems and ($ii$) that the polymer, regarded as a discrete-time stochastic process, is ergodic with decay of correlations. We illustrated the independence of the diffusivity on the polymer correlations by means of a simple model in two different cases. In the first case we produced the polymers by means of a three-states Markov process. Then these polymers were used to build up the potential felt by the particles. In this case we observed that the correlations does not alter the diffusivity and the corresponding effective diffusion coefficient that we obtained from the simulations was consistent with the exact result we derived. In the second case that we analyzed, the polymers were produced by means of the E-M process. The E-M process is known to exhibit long range correlations and the corresponding scaling laws for the correlation functions are known exactly. This feature of the E-M process allowed us to study the diffusivity in a system with a very slow decay of correlations. We could verify the independence on the correlations of the diffusivity; however we found that the effective diffusion coefficient seems to deviated moderately in the case in which the correlations decay as  $C(\ell) \approx \ell^{\theta(0.90)}$ with $\theta (0.90) \approx 0.521$.  We argued that this apparent inconsistence might be due to the fact that the E-M process relaxes slowly to its stationary state. Moreover, we observed that the numerically obtained two-dimensional marginal $v_2(00)$ (corresponding to the realizations of the polymers used for the numerical simulations) exhibits large statistical deviations on every realization of the polymer. This fact clearly affects the diffusivity along every polymer, resulting naturally  in a deviation from the theoretical result. However it is also clear that the diffusion process by itself could be a very slow process since the particles should explore all the polymer in order to have an accurate estimation of the diffusivity. However, due to the existence of large ordered domains in the polymer we could face a very slow convergence of the effective diffusion coefficient. It is therefore necessary a more detailed study of this class of systems particularly focusing in understanding how long is the transient and how does it behave. Knowing these properties could shed some light about the diffusion processes occurring in real systems in which the long-range correlations are naturally present, such as the DNA.

\bigskip

\ack
RSG thanks CONACyT for financial support through Grant No. CB-2012-01-183358. 
CM was supported by the CONICYT-FONDECYT Postdoctoral Grant No. 3140572.

\appendix

\section{Diffusion coefficient in the limit of zero tilt strength}
\label{ape:Deff_zero_tilt}

In Ref.~\cite{salgado2014drift} it has been shown that in the polymer-particle interaction model for diffusion on random media the particle flux can be written exactly in terms of the mean FPT,
\begin{eqnarray}
\label{eq:ape:drift}
J_{\mathrm{eff}} &=& \frac{L}{\langle T_1 \rangle_{\mathrm{p}}},
\end{eqnarray}
where the mean FPT w.r.n. $T_1$ is given by
\begin{equation}
T_1(\mathbf{a}) = \gamma \beta \sum_{m=1}^\infty e^{-m\beta F L} q_{+}(\mathbf{a})q_{-}[\sigma^{-m}(\mathbf{a})] 
+ \gamma \beta I_0(\mathbf{a}).
\label{eq:ape:T1}
\end{equation}
Our goal here is to prove that in the limit of $F=0$ the mobility $\mu$,
\[
\mu = \lim_{F\to 0} \frac{J(F)}{F}
\]
is well defined for any stationary measure $\nu$ defining the polymer ensemble. The only condition we impose is the decay of correlations even at a very slow rate. In other words, we impose a mild condition on the rate of mixing of the dynamical system, which  in turn implies ergodicity~\cite{brin2002intro}. By ergodicity we mean that the time-average of any regular observable $\mathcal{O}: \mathcal{A}^\mathbb{Z} \to \mathbb{R}$, can be interchanged by its space average with respect to the shift-invariant probability measure $\nu$, i.e., 
\[
\int_{ \mathcal{A}^\mathbb{Z}  } \mathcal{O}\,d\nu = \lim_{n\to\infty} \frac{1}{n} \sum_{j=0}^{n-1} \mathcal{O}\circ \sigma^j(\mathbf{a}) \quad \nu\mbox{-almost everywhere}.
\]
Here our observables will be $q_{\pm}$ and $T_{1}$ which are non-negative and for any potential $\psi$ bounded by above, they are also bounded, and furthermore are $\mathrm{L}^{1}(\nu)$, i.e. 
\[
\langle T_1 \rangle_\mathrm{p} = \int_{\mathcal{A}^\mathbb{Z}} T_1 d\nu = \int_{\mathcal{A}^\mathbb{Z}}| T_1 | \,d\nu  < \infty.
\]
Next, we will obtain an expression for the asymptotic behavior of $J_{\mathrm{eff}}/F$ as $F$ approaches to zero. In order to do this, let us first estimate bounds for $\langle T_{1}\rangle_{\mathrm{p}}$. We assume that the system $(\mathcal{A}^{\mathbb{Z}},\sigma, \nu)$ satisfies that for every $f\in \mathrm{L}^{\infty}(\nu)$ and $g\in \mathrm{L}^{1}(\nu)$ there exists a decreasing-to-zero function $r(m)$ such that
\[
\bigg| f\cdot g\circ\sigma^{m}d\nu - \int fd\nu \int g d\nu\bigg|\leq || f ||_{\infty} || g||_{1}\cdot r(m),
\]
where $|| \cdot ||_{\infty}$ and $|| \cdot ||_{1}$ denote the $\mathrm{L}^{\infty}(\nu)$ and $\mathrm{L}^{1}(\nu)$ norms respectively. The rate function $r(m)$ will be precisely given below. 
Let us write $C := || f ||_{\infty} || g ||_{1}$, this is a constant since $q_{\pm}$ are essentially bounded and $L^{1}(\nu)$, and thus we are able to give the following bounds
\begin{eqnarray}
\hspace{-1cm}
\langle T_{1}\rangle_{\mathrm{p}} & \lessgtr &
\gamma\beta \sum_{m=1}^{\infty}e^{-m\beta FL}\left( \int q_{+}d\nu \int q_{-}d\nu \pm Cr(m)\right) +\gamma\beta\int I_{0}d\nu,\\
\hspace{-1cm}
&\lessgtr & \gamma\beta\left( \frac{\langle q_{+}\rangle_{\mathrm{p}}\langle q_{-}\rangle_{\mathrm{p}}}{1-e^{-\beta FL}}\pm C\sum_{m=1}^{\infty}e^{-m\beta FL}r(m) + \int I_{0}d\nu\right).
\end{eqnarray}
Which makes that the particle flux (Eq.~\ref{eq:ape:drift}) be bounded by above and below as follows,
\[
J_{\mathrm{eff}} \lessgtr \frac{L}{\gamma\beta\left( \frac{\langle q_{+}\rangle_{\mathrm{p}}\langle q_{-}\rangle_{\mathrm{p}}}{1-e^{-\beta FL}}\mp C\sum_{m=1}^{\infty}e^{-m\beta FL}r(m) + \int I_{0}d\nu\right)}.
\]
Next, let us consider the approximation as $F\to0$ of the following quantity, 
\[
\lim_{F\to0} F\left( \frac{\langle q_{+}\rangle_{\mathrm{p}}\langle q_{-}\rangle_{\mathrm{p}}}{1-e^{-\beta FL}}\pm C\sum_{m=1}^{\infty}e^{-m\beta FL}r(m) + \int I_{0}d\nu\right).
\]
The third part goes to zero since $\int I_{0}d\nu$ is constant. By limit identities, the first part is given by
\[
\lim_{F\to0} \frac{F\cdot\langle q_{+}\rangle_{\mathrm{p}}\langle q_{-}\rangle_{\mathrm{p}}}{1-e^{-\beta FL}} = \frac{Z_{+}Z_{-}}{\beta L}.
\]
It remains to estimate the summation involving the rate function, for which we make use the assumption on the decay of the correlations. Let us  assume that the rate function $r(m)= m^{-\alpha}$ for any $\alpha>0$. This implies that the sum
\[
\sum_{m=1}^{\infty}e^{-m\beta FL}m^{-\alpha},
\]
can be written as the polylogarithm function~\cite{110}, defined as,
\begin{equation}\label{eq:ape:polylog}
\mathrm{Li}_{\alpha}(x) := \sum_{k=1}^{\infty}\frac{x^{k}}{k^{\alpha}}.
\end{equation}
It is known that the polylogarithm function has the asymptotic behavior~\cite{110}
\begin{equation}\label{eq:ape:approx-poly}
\mbox{Li}_s(e^{-\xi})\approx \Gamma(1-s)\xi^{s-1}
\end{equation}
for $|\xi| \ll 1 $ and for non-integer $s$ (in particular for $0<s<1$), being  $\Gamma$ the well known gamma function. If we consider a $F_0>0$ small enough such that $\beta F_{0}L \ll 1$, for any fixed $\beta$ and $L$, then, one may write for every $F\leq F_{0}$,
\[
F C \sum_{m=1}^{\infty}e^{-m\beta FL} m^{-\alpha} \approx C F\cdot \Gamma(1-\alpha)(\beta F L)^{\alpha -1},
\]
for all $\alpha>0$. In this way we have that
\[
\lim_{F\to0} F C\sum_{1}^{\infty}e^{-m\beta FL}m^{-\alpha} =\lim_{F\to0} C\cdot \Gamma(1-\alpha)(\beta L)^{\alpha -1} F^{\alpha} \to 0,
\]
for all $\alpha>0$ and $\beta, L>0$. Now, using the Einstein relation we see that the effective diffusion coefficient is given by
\[
D_{\mathrm{eff}} = \frac{1}{\beta} \lim_{F\to0}\frac{J_{\mathrm{eff}}}{F} = \frac{1}{\gamma \beta}\frac{L^{2}}{Z_{+}Z_{-}},
\]
which is the expression for the diffusivity anticipated in Eq.~(\ref{eq:Deff_F=0}).
\vspace{2mm}

On one hand, the above treatment gave us the desired expression for the diffusivity when one takes firstly the average on the state space and after that  the approximation when $F$ goes to zero. On the other hand, one would expect that the expression~(\ref{eq:Deff_F=0}) remains valid when one takes first the approximation $F\to 0$ and after that the average with respect to the polymer ensemble. Here we give a result that enables us to argue that this could be actually true. Nevertheless, we left for further research the mathematically rigorous treatment. The importance on this kind of result is that the Einstein relation should hold even if we do not perform the average over the polymer ensemble. This means in particular that the mean FPT $T_1$ should behave as $F^{-1}$ for $F\to 0$ which in turn implies Eq.~(\ref{eq:Deff_F=0}) as we will see below.

For the sake of convenience, we introduce a quantity of interest. Let $N\in\mathbb{N}$ and define the observable $g_{N} : \mathcal{A}^{\mathbb{Z}} \to \mathbb{R}$ as,
\begin{equation}\label{g_N}
g_{N}(\mathbf{a}) := \frac{1}{N}\sum_{m=1}^{\infty}e^{-mz/N}f\circ\sigma^{m}(\mathbf{a}),
\end{equation}
for some $z\in\mathbb{R}^{+}$.  The discussion in the rest of this appendix is to justify the approximation 
\[
\sum_{m=1}^\infty  e^{-mz/N} f\circ \sigma^{m}(\mathbf{a}) \approx  \frac{N}{z} \int_{  \mathcal{A}^\mathbb{Z} } f \, d\mu = \frac{N}{z}\langle f(\mathbf{a})\rangle_\mathrm{p},
\]
for large $N$. 
 
For the moment, let us go to expression for the mean FPT and see that the approximation above may be applied to the summations appearing in $T_1$ in Eq.~(\ref{eq:ape:T1}) if we identify $N$ with $1/F$,
\begin{eqnarray}
T_1(\mathbf{a}) &=& \gamma \beta q_{+}(\mathbf{a}) \sum_{m=1}^\infty e^{-m\beta F L} q_{-}[\sigma^{-m}(\mathbf{a})] 
+ \gamma \beta I_0(\mathbf{a}),
\nonumber
\\
&\approx& \gamma \beta q_{+}(\mathbf{a})\frac{1}{\beta F L} \langle q_{-}(\mathbf{a}) \rangle_{\mathrm{p}} + \gamma \beta I_0(\mathbf{a}).
\end{eqnarray}
Notice that in the limit $F\to 0$ the mean FPT goes to infinity as $F^{-1}$. The term $\gamma \beta I_0(\mathbf{a}) $ appearing in the above expression remains finite if $F=0$, and therefore it can be neglected in further calculations. Then we have that the particle flux behaves asymptotically as
\[
J_{\mathrm{eff} } = \frac{L}{\langle T_1 \rangle_{\mathrm{p}}} \approx \frac{FL^2 }{\gamma} \frac{1}{ \langle q_{+}(\mathbf{a}) \rangle_{\mathrm{p}} \langle q_{-}(\mathbf{a}) \rangle_{\mathrm{p}}},
\]
when $F$ is small enough. Now, using the Einstein relation we see that the effective diffusion coefficient is given by 
\[
D_{\mathrm{eff}} = \frac{1}{\beta} \lim_{F\to 0} \frac{J_{\mathrm{eff}} }{F}   = 
\frac{L^2 }{\gamma \beta} \frac{1}{ Z_{+}Z_{-} }, 
\]
where $Z_{\pm}$ are given by Eq.~(\ref{eq:Zpm})  and which is the expression for the diffusivity anticipated in Eq.~(\ref{eq:Deff_F=0}).

Next, in order to justify our claim, we give the following proposition.

\begin{prop}
Let $(\mathcal{A}^\mathbb{Z}, \sigma, \mathcal{B}, \nu)$ be a measure-preserving dynamical system, where $\sigma : \mathcal{A}^\mathbb{Z} \to \mathcal{A}^\mathbb{Z}$ is the shift mapping, $\mathcal{B} $ is a Borel sigma-algebra on $\mathcal{A}^\mathbb{Z} $  and $\nu$ is a shift-invariant probability measure satisfying decay of correlations at rate $ O(m^{-\alpha})$ for some $\alpha>0$. Let $ f : \mathcal{A}^\mathbb{Z} \to \mathbb{R}$ be a square integrable function on the full shift $\mathcal{A}^\mathbb{Z}$, i.e.,
\[
\int_{  \mathcal{A}^\mathbb{Z} } f^2 d\nu < \infty.
\]
And let $g_{N}$ be the observable defined in~(\ref{g_N}). Then, for every $\epsilon>0$ there exists a sufficiently large $N$ such that
\[
\nu\left(\left\{ \mathbf{a}\in\mathcal{A}^{\mathbb{Z}} : \left| g_{N}(\mathbf{a}) - \frac{1}{z}\int_{\mathcal{A}^{\mathbb{Z}}}f\ d\nu \right| \geq \epsilon \right\} \right) \leq \frac{C\cdot \Gamma(1-\alpha)z^{\alpha -2}}{N^{\alpha}\epsilon^2}.
\]
\end{prop}

This inequality states that the larger we take $N$ the more probable is to find $g_{N}$ around $(1/z)\int fd\nu$ in a $\epsilon$ neighborhood. The proof of this proposition use the decay of the auto-correlation function, for that reason the square integrable assumption on the observable $f$.

\emph{Proof:} 
First, notice that
\begin{eqnarray}
\int_{  \mathcal{A}^\mathbb{Z} } g_{N}\ d\nu &=& \frac{1}{N}\sum_{m=1}^\infty  e^{-mz/N} \int_{ \mathcal{A}^\mathbb{Z}} f\circ \sigma^{m}d\nu,
\nonumber
\\
 &=& \frac{1}{N}\frac{1}{ 1-e^{-z/N}}\int_{\mathcal{A}^{\mathbb{Z}}} f d\nu 
\end{eqnarray}
because $\nu$ is $\sigma$-invariant. For the sake of clarity let us denote by $\bar{f}:= \int_{\mathcal{A}^{\mathbb{Z}}}f d\nu$ and let us define $K : \mathbb{N} \to \mathbb{R}$ as follows,
\[
K(N) :=  \frac{1}{N}\frac{ 1}{ 1-e^{-z/N}} 
\]
and note that for every $z>0$ 
\[
\lim_{N\to\infty} K(N) = \frac{1}{z}.
\]
Observe that for any $\epsilon > 0$ we have 
\begin{eqnarray}
\fl
\nonumber
\nu\left(\bigg\{ \left|g_N(\mathbf{a}) -  \frac{\bar{f}}{z} \right| \geq \epsilon\bigg\} \right) 
&=& \nu\left(\bigg\{ \left|g_N(\mathbf{a})  - K(N) \bar{f} +K(N) \bar{f} - \frac{\bar{f}}{z} \right| \geq \epsilon\bigg\} \right),\\
\nonumber
&\leq& \nu\left(| g_{N} - K(N)\bar{f} | >\epsilon/2\right) + \nu \left( | K(N)\bar{f} - \bar{f}/z | > \epsilon/2 \right).
\end{eqnarray} 
For the second summand in the right-hand side of the inequality, there is no probability. So, there exists a $N^{*}(\epsilon)$ such that for all $N\geq N^{*}$, $ | K(N)\overline{f} - \overline{f}/z | \leq \epsilon/2$, and so that part eventually vanishes. While for the first summand, by the Chebyshev inequality we have that for all $N\in \mathbb{N}$,
\[
\nu\left(\bigg\{ \left| g_N(\mathbf{a}) -  K(N)\bar{f} \right| \geq \epsilon/2 \bigg\} \right) 
\leq 
\frac{4 \mbox{Var} (g_N)}{\epsilon^2}.
\]
The variance of $g_N$ can be rewritten in terms of the auto-correlation function,
\[
\mbox{Var} (g_N)  = \int g_N^2 d\nu - \left( \int g_Nd\nu  \right)^2 = \int g_N^2d\nu  - K^2(N) \bar{f}^2.
\]
We need to calculate $ \int g_N^2 d\nu$ and we obtain that
\[
\int g_N^2  d\nu = \frac{1}{N^2} \sum_{m=1}^\infty \sum_{k=1}^\infty e^{-mz/N} e^{-kz/N} \int \left( f\circ \sigma^m\right) \left(  f\circ \sigma^k\right) d\nu,
\]
using the $\sigma$-invariance of $\nu$ we get,
\begin{eqnarray}
\hspace{-2.3cm}
\mbox{Var} (g_N)  &=&  \frac{1}{N^2} \sum_{m=1}^\infty \sum_{k=1}^\infty e^{-mz/N} e^{-kz/N} \Bigg[   \int f \cdot f \circ \sigma^{m-k} \, d\nu - \left( \int f \, d\nu \right)^2\Bigg]
\nonumber 
\\
\hspace{-2.3cm}
&=&
\frac{1}{N^2} \sum_{m=1}^\infty \sum_{k=1}^\infty e^{-mz/N} e^{-kz/N} C_f(m-k).
\label{eq:Var_G_sum1}
\end{eqnarray}
where $C_f$ stands for the auto-correlation function of the observable $f$, i.e.,
\[
C_f (\ell) :=  \int f \cdot f \circ \sigma^{\ell} \, d\nu - \left( \int f \, d\nu \right)^2.
\]
The expression for the variance of $g_N$ given in Eq.~(\ref{eq:Var_G_sum1}) can rewritten by changing the summation order, as follows
\begin{eqnarray*}
\hspace{-2cm}
\mbox{Var} (g_N)  &=&  \frac{2}{N^2} \sum_{\ell = 1}^\infty \sum_{n=1}^\infty e^{-2n z/N} e^{-\ell z/N} C_f(\ell) 
+  \frac{1}{N^2} C_f(0) \sum_{n=1}^\infty e^{-2n z/N}.
\nonumber 
\\
\hspace{-2cm}
 &=&  K(N/2) \frac{1}{N} \sum_{\ell = 1}^\infty e^{-\ell z/N} C_f(\ell) 
+  \frac{1}{N} C_f(0) \frac{K(N/2)}{2}.
\end{eqnarray*}
Now, the assumption on the decay of correlations becomes cricial. Let us assume that the auto-correlation function for $f$ behaves as $C_f (\ell) = C_0 \ell^{-\alpha}$ for any $\alpha >0$ and some $C_0 >0$. In this case we can write the variance of $g_N$ as follows
\begin{eqnarray}
\nonumber
\mbox{Var} (g_N)  &=&  K(N/2) \frac{C_0}{N} \sum_{\ell = 1}^\infty \ell^{-\alpha} e^{-\ell z/N} 
+  \frac{1}{N} C_f(0) \frac{K(N/2)}{2}
\\
\nonumber
&=&  K(N/2) \frac{C_0}{N} \mbox{Li}_\alpha (e^{-z/N}) 
+  \frac{1}{N} C_f(0) \frac{K(N/2)}{2}.
\end{eqnarray}
In the second equation we have rewritten the expression for the variance of $g_{N}$ using the polylogarithm function~(\ref{eq:ape:polylog}).
From the approximation~(\ref{eq:ape:approx-poly}) we are able to write down an asymptotic expression for the variance of $g_N$ for a sufficiently large $N$,
\[
\mbox{Var} (g_N)  \approx  K(N/2) \frac{C_0}{N}\Gamma(1-\alpha)\left(\frac{z}{N}\right)^{\alpha-1} 
+  \frac{1}{N} C_f(0) \frac{K(N/2)}{2}.
\]
Finally, since $K(N/2) \to 1/z$ as $N\to \infty$, we easily see that $ \mbox{Var} (g_N)  \to 0$ in the same limit, and specifically
\[
\mbox{Var} (g_N)  \approx C_0 \Gamma(1-\alpha) z^{\alpha - 2} N^{-\alpha}.
\]
The latter implies that we can chose an appropriate constant $C$ such that for a sufficiently large $N$
\begin{eqnarray}
\nu\left(\bigg\{ \left|g_N(\mathbf{a}) -  \frac{\bar{f}}{z} \right| \geq \epsilon\bigg\} \right) 
&\leq& 
\frac{ C \cdot \Gamma(1-\alpha) z^{\alpha - 2}}{N^{\alpha}\epsilon^2}.
\nonumber
\end{eqnarray} 
Which concludes the proof the proposition.

\hfill\ensuremath{\square}
\vspace{2mm}

\section{The $n$-dimensional marginals for the expansion-modification system}
\label{ape:marginals_EM}

The 1-dimensional marginals for the E-M system can be understood as the relative frequency of occurrence of symbols (monomers) of every kind along a typical realization of the symbolic sequence (the polymer) generated by the E-M process at the stationary state. For the sake of simplicity here we will call ``polymer'' to an infinite symbolic sequence (i.e., a point in the space $\{0,1\}^\mathbb{Z}$).  

The processes generating  the infinitely long polymer of the E-M system, give us the way to obtain the marginals as follows.  First, let $\mathbf{a}_0\in \{0,1\}^\mathbb{Z}$ be an ``initial polymer'' (of length $0$) and allow it evolve, coordinatewise, according to the E-M processes defined as
\[
x \mapsto \left\{\begin{array}{cl} \overline{x} &  \mbox{ with probability } 1-p \\ 
                                   x x & \mbox{ with probability } p.
                  \end{array}\right.
\]
These rules generate a sequence of ``polymers'' of finite size  $\mathbf{a}_0 \mapsto \mathbf{a}_1 \mapsto \mathbf{a}_2  \cdots$ that attain a stationary state for $t\to\infty$. The E-M system is actually a discrete-time Markov process~\cite{salgado2013exact}. Let us call $f_0 (t)$ and $f_1(t)$ the fraction of 0's and 1's in the symbolic chain $\mathbf{a}_t$ at time $t\in \mathbb{N}$. This means that in any subsequence of $\mathbf{a}_t$ of $n$ symbols we will find typically $n f_0 (t) $ zeroes and $nf_1 (t)$ ones. We will obtain a recurrence relation for $f_0$ and $f_1$.  First notice that the chain $\mathbf{a}_{t+1}$ is generated from $\mathbf{a}_t$ by means of a fraction of $p$ expansions and $1-p$ modifications. This means that a subsequence $\mathbf{b}$ of size $n$ in $\mathbf{a}_{t}$  generate a sequence of size $n(1+p)$ in $\mathbf{a}_{t+1}$. The total symbols that are expanded in $\mathbf{b}$ is $pn$ and the total symbols that are modified is  $(1-p)n$. From the total symbols that are ``expanded'', a fraction $f_0$ of them are zeroes and a fraction $f_1$ of them are ones. Analogously from the total symbols that are ``modified'', a fraction $f_0$ of them are zeroes and a fraction $f_1$ of them are ones. This means that the number of symbols that are zeroes  in the new chain  has two contributions: the number of 0's that were expanded $2npf_0(t)$ plus the number of 1's that were modified $n(1-p) f_1(t)$ becoming zeroes in the new chain. An analogous statement follows immediately for $f_1(t)$ . Since the size of the new chain is $(1+p)n$ we have that the fractions $f_0(t)$ and $f_1 (t)$ obey the following recurrence relations,
\begin{eqnarray}
f_0 (t+1) &=&\frac{ 2p f_0(t) + (1-p) f_1(t)}{1+p}
\\
f_1 (t+1) &=& \frac{ (1-p) f_0(t) + 2p f_1(t)}{1+p}.
\end{eqnarray}
Notice that this defines a one-step Markov chain with probability matrix $M_1 :  \{0,1\}\times \{0,1\} \to [0,1] $,
\begin{eqnarray}
M_1 &=& \frac{1}{1+p} \left(
  \begin{array}{cc}
   2p & 1-p  \\
   1-p & 2p   
   \end{array} \right).
  \label{eq:ape:M1}
\end{eqnarray}
In the limit of $t\to \infty$ the fractions $f_0$ and $f_1$ tend to the unique invariant probability vector for $M_1$. It can be seen that such a probability vector is given by
\begin{eqnarray}
v_1 &=& \left(\frac{1}{2},\frac{1}{2}\right).
\label{eq:ape:v1}
\end{eqnarray}

Next we can generalize the concept of fractions of zeroes and ones, to the \emph{relative frequency} $f_t(\mathbf{w})$ with which a word $\mathbf{w} \in \{0,1\}^k$ of size $k$ occurs along a the polymer $\mathbf{a}_t \in  \{0,1\}^\mathbf{Z}$ at time $t$. 
\begin{equation}
f_t(\mathbf{w}) := \lim_{n\to\infty} \sum_{j=0}^{n-1} \mathbb{I}_ {[\mathbf{w}]}\left(\sigma^j (\mathbf{a}_t) \right)
\end{equation}
where $\mathbb{I}$ is the indicating function, which takes the value $1$ if $\mathbf{a}_t$ belongs to the set $[\mathbf{w}]$ and take the value $0$ if $\mathbf{a}_t$ does not belong to the set $[\mathbf{w}]$. The set $[\mathbf{w}]$ called \emph{cylinder} is defined as the set of all infinite sequences $\mathbf{x} \in \{0,1\}^\mathbb{Z}$ with prefix $\mathbf{w}$, i.e., 
\[
[\mathbf{w}] := \bigg\{ \mathbf{x} \in \{0,1\}^\mathbb{Z} \, : \, x_0 = w_0, x_1 = w_1, \dots, x_{k-1} = w_{k-1}  \bigg\}.
\]
In other words, we calculate the frequency that a word of size $k$ appears in a polymer by counting the number times such a word appears in a ``window'' of size $k$ which slides along the polymer.  Let us analyze the case $k=2$. To obtain the stationary 2-dimensional marginals, or equivalently, the stationary relative frequencies of words of size $2$, we can obtain recurrence relations as in the 1-dimensional case. First notice that a  word of size $2$ generate words of sizes $2$, $3$ and $4$ by means of expansions and modifications. We have four cases that should be considered. For example take the word $\mathbf{a} = 00$. This word can be transformed under the E-M process into $\mathbf{b} = 0000, 100, 001,11$. Each case corresponds to one of the four possibilities of the process: $\mathtt{ee}$, $\mathtt{em}$, $\mathtt{me}$ and  $\mathtt{mm}$ ($\mathtt{e}$ stands for ``expansion'' and $\mathtt{m}$ stands for ``modification''). First notice that in the case of two expansions ($00 \stackrel{\mathtt{ee}}{\to} 0000$) we obtain, by sliding a window of size two, three times the word 00. However, this fact does not mean that pure expansions has the effect of triplicate the frequency of appearance of $00$. For example, the block $00000$ transforms under pure expansion into the block $0000000000$. A direct counting shows, by sliding a window of size two, that the word $00$ appears four times before the transformation and nine times after such a process. In general, the number of times that the word $00$ appears in the block resulting from pure expansions  of a block consisting of $n$ zeros, is $2n +1$. This means that the frequency of appearance of $00$ is duplicated in the limit of an infinite polymer. In other words, the effective number of words that produces the word 00 under the process $\mathtt{ee}$ is two and are $00$, $00$. A similar analysis shows that, in order to count correctly the number of times a word appears under the processes $\mathtt{em}$, $\mathtt{me}$, and $\mathtt{mm}$ we should take into account certain ``rules'' which are summarized in Table~\ref{tab:transformations}.
\Table{\label{tab:transformations}
Effective number of words of size two given by the expansion-modification process. This rules allows calculate the recurrence relations for the evolution of frequency of appearance of words of size two.  
}
\br
 & & \centre{4}{Initial words}\\
\ns
& & \crule{4}\\
Process &Probability            & $00$ gives  & $01$ gives  & $10$ gives   & $11$ gives \\
\mr
$\mathtt{ee}$ & $p^2$         & $00$, $00$ & $00$, $01$  & $11$, $10$  & $11$, $11$  \\
$\mathtt{em}$ & $p(1-p)$    & $00$, $01$ & $00$, $00$  & $11$, $11$  & $11$, $10$  \\
$\mathtt{me}$ & $p(1-p)$    & $10$           & $11$            & $00$            & $01$            \\
$\mathtt{mm}$ & $(1-p)^2$ & $11$           & $10$            & $01$            & $00$            \\
\br
\end{tabular}
\end{indented}
\end{table}
With these rules it is easy to see that the relative frequencies with which the words $00$, $01$, $10$ and $11$ appear after a process of expansions and modifications of an infinite polymer, are given by,
\begin{eqnarray}
\fl
f_{t+1} (00)  &=& \frac{1}{1+p}\bigg( (2 p^2 +pq ) f_t(00) + (p^2 + 2 pq) f_t (01) + pq f_t (10) +q^2 f_t (11)  \bigg)
\nonumber 
\\
\fl
f_{t+1} (01)  &=& \frac{1}{1+p}\bigg( pq f_t(00) + p^2 f_t (01) + q^2 f_t (10) + pq f_t (11)  \bigg)
\nonumber 
\\
\fl
f_{t+1} (10)  &=& \frac{1}{1+p}\bigg( pq f_t(00) + q^2 f_t (01) + p^2  f_t (10) +pq f_t (11)  \bigg)
\nonumber 
\\
\fl
f_{t+1} (11)  &=& \frac{1}{1+p}\bigg( q^2 f_t(00) + pq f_t (01) + (p^2 + 2pq) f_t (10) + (2p^2 + pq) f_t (11)  \bigg).
\label{eq:recurrence_f2}
\end{eqnarray}
The above recurrence relations can be recast into a more compact form by defining the probability vector $p_2$  (the time-dependent two-dimensional marginals) whose components are the relative frequencies at time $t$,
\[
p_2 (t) := \bigg( f_{t+1} (00), \, f_{t+1} (01),\, f_{t+1} (10),\, f_{t+1} (11) \bigg),
\] 
in terms of which we rewrite Eq.~(\ref{eq:recurrence_f2}) as a matrix equation,
\[
p_2 (t+1) = p_2(t) M_2
\]
where the matrix $M_2$ defined as 
\begin{eqnarray}
M_2 &=& \frac{1}{1+p} \left(
  \begin{array}{cccc}
   2p^2+pq  & pq   & pq   & q^2 \\
   p^2 + 2pq & p^2  & q^2  & pq  \\
   pq       & q^2  & p^2  & p^2 +2pq \\
   q^2      & pq   & pq   & 2p^2+pq     
  \end{array} \right).
  \label{eq:ape:M2}
\end{eqnarray}
In the limit of $t\to \infty$, it has been proved in Ref~\cite{salgado2013exact} that the system reaches a stationary state. This means that the time-dependent probability vector $p_2 (t)$ is well defined in the limit $t \to \infty $. The probability vector at infinity is the left eigenvector of $M_2$ , which we denote by  $v_2$, associated to the (largest) eigenvalue $\lambda = 1$, 
\[
v_2 = \lim_{t\to\infty} p_2(t),
\]  
and $v_2$ satisfies the equation
\[
v_2 = v_2 M_2,
\]
through which we calculated the $2$-dimensional marginal given in Eq~(\ref{eq:v2}).

\bigskip
\section*{References}

\nocite{*}

\bibliography{EffDiffDisorderedPotentials_ref.bib}

%
%
%
%
%
%

\end{document}